\documentclass[11pt,a4paper]{article}
\usepackage{jheppub}
\usepackage{graphicx}
\title{A scattering approach to some aspects of the Schwarzschild Black Hole}
\author{Bernard Raffaelli}
\note{On leave from September 2012}
\affiliation{University of Nice,\\
Facult\'e des Sciences, Parc Valrose\\ 28, avenue Valrose, 06108 Nice Cedex 2,FRANCE}
\emailAdd{bernard.raffaelli@unice.fr}

\abstract{In this paper, we consider a massless field, with spin $j$, in interaction with a Schwarzschild black hole in four dimensions, focusing mainly our study on the $s$-wave scattering. First, using a Fourier analysis, we show that one can have a simple and natural description of the Physics near the event horizon without using any conformal field approaches. Then, within the same ``scattering picture'', we derive analytically the imaginary part of the highly damped quasinormal complex frequencies and, as a natural consequence of our analysis, we show that thermal effects and in particular Hawking radiation, can be understood through the scattering of an ingoing $s$-wave by the non null barrier of the Regge-Wheeler potential associated with the Schwarzschild black hole. Finally, with the help of the well-known expression of the highly damped quasinormal complex frequencies, we propose a heuristic extension of the ``tripled Pauli statistics'' suggested by Motl, some years ago.}
\begin{document}
\maketitle

\section{Introduction}
Scattering by black holes (BH) have been studied intensively since the pioneer work of Matzner, in 1968 \cite{Matzner1968}. There have been a lot of works using different approaches. One of them uses the complex angular momentum (CAM) theory, the $S$-matrix formalism and the associated Regge pole techniques. In this framework, the authors have shown ~\cite{YDAFBJ2003,YDAFBR2010} that the weakly damped quasinormal mode (QNM) complex frequencies of a wide class of static, asymptotically flat and spherically symmetric BH of arbitrary dimension, can be understood as Breit-Wigner type resonances generated by the interferences and damping of a family of ``surface waves'' lying close to their photon sphere, whose existence is of fundamental importance. It should be noted that, in~\cite{YDAF2009}, they have shown that this ``surface waves'' interpretation can also be applied to the BTZ BH, a non-asymptotically flat spacetime, i.e. in a framework where the notion of an $S$-matrix does not exist, by extending a powerful formalism introduced several decades ago by Sommerfeld~\cite{Sommerfeld49}. Finally, by noting that each ``surface wave'' is associated with a Regge pole of the corresponding $S$-matrix, this approach permits to construct analytically the spectrum of the weakly damped QNM complex frequencies of the corresponding BH, beyond the leading order term. Physically, it gives a powerful and quite elegant way to obtain a semiclassical interpretation of BH resonance phenomena. However, even if they have computed the absorption cross section (beyond the eikonal order) and the greybody factor in such scheme, the WKB approximation used to obtain analytically the expression of the Regge poles restricts the study to ``high frequency'' regime, i.e. for frequencies of order $\omega \gg (2M)^{-1}$, where $M$ is the mass of the Schwarzschild BH and thus limits the analysis to a scattering description of weakly damped QNM, formally associated to high angular momenta. Thus, highly damped QNM, Hawking radiation, and more generally, the physics near the BH horizon seems to be out of such framework. Nevertheless, in this paper, we claim that one could have a physical description of the near horizon physics without using any conformal field theories but by focusing, precisely, on the scattering of an ingoing $s$-wave by the non null barrier of the Regge-Wheeler potential of the Schwarzschild BH. In section \ref{generalities}, we introduce some notations, assumptions and well-known results linked to a scattering analysis and the $S$-matrix formalism. In section \ref{NHL}, after having discussed the ``Rindler approximation'', we will compute some results which have been assumed or derived by Solodukhin but within a conformal field framework \cite{Solodukhin2004}. In section \ref{HDQNM}, we will focus on the ``far horizon limit'' and we will show that one can easily compute the relevant physical quantities needed for a scattering analysis. They will allow us to compute the exact expression of the imaginary part of highly damped QNM complex frequencies. In section \ref{scattering_statistics}, using the results obtained in section \ref{HDQNM}, we will derive explicitly the reflection coefficient related to the scattering of the ingoing $s$-wave by the non null barrier of the Regge-Wheeler potential of the BH which is exactly the Boltzmann weight associated with the Hawking radiation, giving, in a purely scattering picture, another possible explanation of BH thermal effects. Finally, in section \ref{statistics}, extending some reflexions due to Motl \cite{Motl2002}, we will claim by an heuristic approach that the real part of these highly damped QNM complex frequencies is intimately linked to some ``exotic'' statistics, going beyond the Bose-Einstein and the Fermi-Dirac ones, which could be expected near any BH horizon where a theory of quantum gravity is maybe unavoidable.

\section{Generalities and notations}\label{generalities}

\begin{figure}[!t]
\centering
\includegraphics[scale=1]{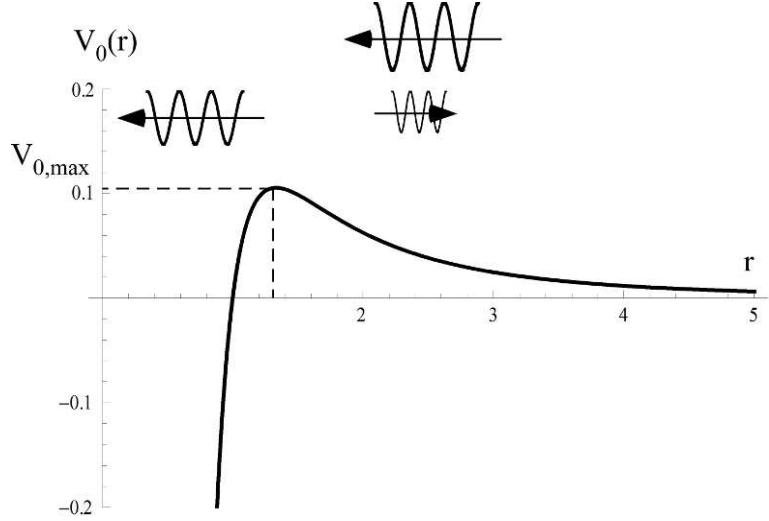}
\caption{``$\ell = 0$'' Regge-Wheeler potential; For the Schwarzschild BH ($2M=1$), the maximum of the Regge-Wheeler potential is evaluated at $V_{0,\text{max}}=(27/256)\, (2M)^{-2}\approx 0.11\, (2M)^{-2}$ and located near the event horizon, i.e. at $r=(4/3)\, (2M)$}
\label{0RWPot}
\end{figure}

We consider first a static spherically symmetric four-dimensional spacetime with metric
\begin{equation} \label{metric_BH}
ds^2=-f(r)dt^2+\frac{dr^2}{f(r)}+r^2d\sigma^{2}.
\end{equation}
Here $d\sigma^{2}$ denotes the line element on the unit sphere $S^{2}$. It should be noted that a metric such as (\ref{metric_BH}) does not describe the most general static spherically symmetric spacetime but it will permit us to easily apply the following results to the Schwarzschild case.\\
In Eq.~(\ref{metric_BH}), we shall assume that $f(r)$ is a function of the usual radial Schwarzschild coordinate $r$ satisfying the properties:

\begin{itemize}
\item (i) There exists an interval $I=]r_h,+\infty[ \subset \bf{R}$ with $r_h>0$ such as $f(r)>0$ for $r \in I$.
\item (ii) $r_h$ is a simple root of $f(r)$, i.e.
\begin{equation}\label{Assump_f_1}
f(r_h)=0 \qquad \text{and} \qquad f'(r_h)\neq 0 \qquad \text{and} \qquad f^{(2)}(r_h)>0.
\end{equation}
\end{itemize}
These assumptions indicate that the spacetime considered has an single event horizon at $r_h$, its exterior corresponding to $r \in I$. If we want to work with an asymptotically flat spacetime, in order to define an $S$-matrix, we need to impose the condition
\begin{equation}\label{Assump_f_2}
\underset{r \to +\infty}{\lim}f(r)=1.
\end{equation}
The Klein-Gordon wave equation for a massless scalar field propagating on a general gravitational background is given by
\begin{equation}\label{WaveEq}
\Box \Phi=g^{\mu\nu}\nabla_{\mu}\nabla_{\nu}\Phi=
\frac{1}{\sqrt{-g}}\partial_{\mu}\left(\sqrt{-g}g^{\mu\nu}\partial_{\mu}\Phi\right)=0.
\end{equation}
If the spacetime metric is defined by (\ref{metric_BH}), after separation of variables and introduction of the radial partial wave
functions $\Phi_\ell(r)$ with $\ell=0,1,2, \dots$, this wave equation reduces to the Regge-Wheeler equation
\begin{equation}\label{RW}
\frac{d^2 \Phi_\ell}{dr_*^2} + \left[ \omega^2 - V_\ell(r)\right]\Phi_\ell=0.
\end{equation}
Here we have assumed a harmonic time dependence $\exp(-i\omega t)$ for the massless scalar field. The variable $r_\ast=r_\ast(r)$ is the well-known tortoise coordinate defined, for $r \in I$, by the relation $dr_\ast/dr=1/f(r)$. Moreover, it is worth noting that the previous definition of $r_\ast$ provides a bijection $r_\ast=r_\ast(r)$ from $I$ to $]-\infty,+\infty[$.\\
In Eq.~(\ref{RW}), $V_\ell(r)$ is the Regge-Wheeler potential associated to the massless scalar field.
\begin{equation}\label{RWPotscalar}
V_\ell(r)=f(r) \left[ \frac{\ell(\ell+1)}{r^2}+\frac{1}{r}f'(r)\right].
\end{equation}
In four dimensions, we recall the well-known cases of non zero spin $j$ fields propagating on the BH background described by (\ref{metric_BH}). Indeed, one can be interested, for example, in the propagation of an electromagnetic test-field $(j = 1)$ or a linearized perturbation of the metric $(j = 2)$. Thus, for more general situations in four dimensions, the Regge-Wheeler potential can be written as
\begin{equation}\label{RWPot}
V_\ell(r)=f(r) \left[ \frac{\ell(\ell+1)}{r^2}+\frac{J}{r}f'(r)\right]
\end{equation}
where $J=1-j^2$, with $j$ the spin of the massless field under consideration. We recall that this result is valid when the BH does not carry any electric or magnetic charge. However, we would like to refer the reader to \cite{YDAFBR2010} for a $d$-dimensional generalization of the scattering of a massless scalar field by a static and spherically symmetric BH through the semi-classical CAM techniques.
\newline
According to (\ref{RWPot}), it should be noted that $\lim_{r \to r_h} V_\ell(r)=0$ and $\lim_{r \to +\infty} V_\ell(r)=0$ and therefore the solutions of the radial equation (\ref{RW}) have a $\exp(\pm i \omega r_{\ast})$ behavior at the horizon and at infinity. In other words, for a given angular momentum index $\ell$, a general solution of the Regge-Wheeler equation (\ref{RW}), satisfies the following asymptotic behaviors:
\begin{equation}\label{bc1}
\Phi_{\omega,\ell} (r_\ast) \underset{r_\ast \to -\infty}{\sim} e^{-i\omega r_\ast }
\end{equation}
which is a purely ingoing wave at the event horizon and which has the following general expression at spatial infinity $r_\ast \to +\infty$
\begin{equation}\label{bc2}
\Phi_{\omega,\ell}(r_\ast)\underset{r_\ast \to +\infty}{\sim} A_{in}e^{-i\omega r_\ast}+A_{out}e^{+i\omega r_\ast}.
\end{equation}
Moreover, by considering the wronskian of two linearly independent solution of Eq.~(\ref{RW}) at $r_\ast=\pm\infty$, we obtain
\begin{equation}\label{Wrsk}
1+\left|A_{out}\right|^2=\left|A_{in}\right|^2.
\end{equation}
As independent solutions we have chosen Eq.~(\ref{bc2}) and its complex conjugate. Furthermore, introducing the transmission and reflection amplitudes $T_\ell$ and $R_\ell$ defined by
\begin{subequations}
\begin{eqnarray}
&&T_\ell=\frac{1}{A_{in}}\\
&&R_\ell=\frac{A_{out}}{A_{in}}
\end{eqnarray}
\end{subequations}
we can rewrite Eq.~(\ref{Wrsk}) in the more familiar form
\begin{equation}
\left|T_\ell\right|^2+\left|R_\ell\right|^2=1.
\end{equation}
Thus, from Eq.~(\ref{bc2}), a QNM, which is defined as a purely ingoing wave at the event horizon and a purely outgoing wave at infinity, corresponds to $A_{in}=0$. Therefore, the part of the ingoing wave that will not be absorbed by the BH will be reflected back to spatial infinity. It should be noted that these quantities can be deduced from the $S$-matrix elements. Here, our problem is spherically symmetric so, the $S$-matrix is diagonal. We recall that the $S$-matrix elements, noted $S_\ell(\omega)$, are related to the reflection amplitude by
\begin{equation}  
S_\ell(\omega)=(-1)^{\ell+1}R_\ell(\omega).
\end{equation}
It has been shown in \cite{YDAFBJ2003,YDAF2009,YDAFBR2010} (and references therein) that the $S$-matrix permits to analyze the resonant aspects of the considered BH as well as to construct the form factor describing the scattering of a monochromatic scalar wave. In this paper, we are mainly interested in the scattering of an ingoing $s$-wave, i.e. $\ell=0$, for which one has, on one hand
\begin{equation}  
S_0(\omega)=-R_0(\omega),
\end{equation}
and on the other hand, 
\begin{equation}
V_0(r)=\frac{J}{r}f'(r)f(r).
\end{equation}
For a Schwarzschild BH of mass $M$, i.e. $f(r)=1-2M/r$, and a massless scalar field ($j=0$), the Regge-Wheeler has a maximum $V_{0,\text{max}}$ for $r=(4/3)\,r_h$ \cite{Matzner1968}. As illustrated in Fig.~\ref{0RWPot}, we will show in section \ref{HDQNM} that the non null potential barrier, for which we will associate an energy $\omega_\text{min}=\sqrt{V_{0,\text{max}}}$, could be at the origin of the Hawking radiation and more generally of thermal effects.

\section{The near horizon limit}\label{NHL}

\subsection{The Rindler approximation}
In order to deal with the spacetime structure in the very vicinity of the BH horizon, i.e. the ``near horizon limit'', we first naturally expand the function $f(r)$ around the value $r=r_h$.
\begin{equation}\label{taylor_f}
f(r) \sim f'(r_h)(r-r_h)+\underset{r \to r_h}{\cal O}(r-r_h)^2.
\end{equation}
Then inserting (\ref{taylor_f}) into (\ref{metric_BH}), we obtain
\begin{equation}
ds^2=-f'_h(r-r_h)dt^2+\frac{dr^2}{f'_h(r-r_h)}+r_h^2d\Omega^{2}.
\end{equation}
We introduce the new variable $\rho$ defined by
\begin{equation}
d\rho=\frac{dr}{\sqrt{f'_h(r-r_h)}}
\end{equation}
such as
\begin{equation}
\rho=2\sqrt{\frac{r-r_h}{f'_h}} \qquad \text{or} \qquad r-r_h=\frac{f'_h}{4}\rho^2.
\end{equation}
Thus, the metric reads
\begin{equation}\label{rindler_approx}
ds^2=-\kappa^2 \rho^2 dt^2+d\rho^2+r_h^2 d\Omega^{2}
\end{equation}
which is regular at $\rho=0$ and where $\kappa=(1/2)f'_h$ is the well-known surface gravity of the Killing horizon at $r=r_h$.
This ``near horizon limit'' form of the metric is of course a ``Rindler approximation'' with a corresponding constant acceleration $\kappa$. From Eq.~(\ref{taylor_f}), we can obtain the behavior of the tortoise coordinate $r_\ast$ near the event horizon
\begin{equation}
r_\ast=\int \frac{dr}{f(r)} \sim \frac{1}{f'_h} \int \frac{dr}{(r-r_h)}
\end{equation}
which gives
\begin{equation}\label{tort1}
r_\ast(r)=\frac{1}{f'_h}\ln\left(\frac{r}{r_h}-1\right),
\end{equation}
where the constant of integration is chosen such as $r=2r_h$ implies $r_\ast=0$. From (\ref{tort1}), we write the usual radial coordinate $r$ as a function of $r_\ast$ 
\begin{equation}
r-r_h=r_h \exp\left(f'_h r_\ast\right)
\end{equation}
which allows us to write Eq.~(\ref{taylor_f}) as
\begin{equation}\label{taylor_ftort}
f(r_\ast) \sim r_h\, f'_h \exp\left(f'_h r_\ast\right)+\frac{1}{2}r_h^2\,f^{(2)}_h \exp\left(2f'_h r_\ast\right)+\underset{r_\ast \to 0}{\cal O}\left(\exp\left(3f'_h r_\ast\right)\right),
\end{equation}
where $f_h^{(p)}=(d^pf/dr^p)|_{r_h}$. Then one can express the Regge-Wheeler equation in the ``near horizon limit'' as it has been done in \cite{Solodukhin2004}, but, as we will see later, it is of no importance in this paper. In terms of the tortoise coordinate, the metric (\ref{metric_BH}) for the Schwarzschild geometry, i.e. $f(r)=1-2M/r$, in the near horizon limit is given by
\begin{equation}\label{metric_tortoise}
ds^2=e^{2\kappa r_\ast}\left(-dt^2+dr_\ast^2\right)+r_h^2d\sigma^2
\end{equation}
where $\kappa=1/4M$ and $r_h=2M$. Of course, if one chooses a hyperplane such as $d\sigma=0$, then in the $(t, r_\ast)$ coordinates, the metric is conformally flat.\\
For the following, we introduce the ``null tortoise coordinates'' defined by
\begin{subequations}\label{uv}
\begin{eqnarray}
&&u=t+r_\ast\\
&&v=t-r_\ast
\end{eqnarray}
\end{subequations}
and the associated Kruskal coordinates
\begin{subequations}\label{UV}
\begin{eqnarray}
&&\kappa U=e^{\kappa u}\\
&&\kappa V=-e^{-\kappa v}.
\end{eqnarray}
\end{subequations}

\subsection{The time dependent ``Doppler-gravitational'' shift effect}

Now, let us consider, near the Schwarzschild event horizon, a ``local inertial frame'' with coordinates $(T,R)$, i.e. a timelike coordinate $T$ and a spacelike coordinate $R$, such as the Kruskal-Szekeres coordinates are defined as ``null coordinates'':
\begin{eqnarray}
&&U=T+R\\
&&V=T-R.
\end{eqnarray}
We recall that the $(U,V)$ Kruskal coordinate system is analogous to the Rindler coordinates while the $(T,R)$ coordinate system is analogous to the Minkowski one. Moreover, they cover the entire spacetime manifold of the maximally extended Schwarzschild solution, being well-behaved everywhere outside the physical singularity at $r=0$.
\newline
Furthermore, since we are working with an $s$-wave which is spherically symmetric, one could always approximate it locally as a ``plane wave'' propagating in the radial direction. So, in the previously defined ``local inertial frame'' $(T,R)$, we can consider a locally (monochromatic) ``plane wave'' propagating towards the BH. Up to an amplitude coefficient, it could be written as
\begin{equation}
\Phi_\text{inertial}(T,R) \propto \exp\left[-i\omega(T+R)\right].
\end{equation}
Thus, according to eqs.~(\ref{uv}) and (\ref{UV}), the locally (monochromatic) ingoing plane wave $\Phi_\text{inertial}(T,R)$ becomes
\begin{equation}
\Phi(t,r_\ast)=\exp\left[-i\left(\frac{\omega}{\kappa}\right)e^{\kappa(t+r_\ast)}\right]
\end{equation}
in the $(t,r_\ast)$ coordinates where it is obviously not a monochromatic plane wave. It is worth noting that keeping the variable $u=t+r_\ast$, the field reads
\begin{equation}\label{horizon_state}
\Phi(u)=\exp\left[-i\left(\frac{\omega}{\kappa}\right)e^{\kappa u}\right]
\end{equation}
which is analogous to the ``horizon state'' $\phi_H$ introduced by Solodukhin \cite{Solodukhin2004} in the framework of a conformal field theory, if one identifies $\omega/\kappa$ with the parameter $\mu_H$. Physically, the ingoing wave $\Phi(u)$ is not monochromatic simply because of the time dependent Doppler shift effect due to the gravitational field of the Schwarzschild BH. 
\newline
According to the previous relations, one could conclude that an observer looking at the ingoing wave actually sees a superposition of ``plane waves'' (cf. \cite{Alsing_Milonni2004}). Indeed, if we write the inverse Fourier transform, we have
\begin{equation}\label{inverseTF}
\Phi(u)=\int_{-\infty}^{+\infty}d\Omega\,\hat{\Phi}_0(\Omega)\,e^{-i\Omega u}.
\end{equation}
The expression of $\Phi(u)$ can be understood as resulting from a superposition of ingoing and outgoing ``plane waves'' near the Schwarzschild event horizon where $\hat{\Phi}_0(\Omega)$ is the amplitude of the ``plane wave'' of frequency $\Omega$. In other words, we focused first on a single plane wave, i.e. a single frequency. We found that in the freely falling frame it becomes a superposition of ``plane waves''. Now, if we want to quantize the considered spin $j$ field, we have to identify each plane wave with single particle states. In \cite{Solodukhin2004}, Solodukhin talks about a transition between the ``horizon state'' and the outgoing propagating wave. Here, the interpretation from the scattering point of view is analogous and far more natural. There is a transition from a single ``plane wave'', i.e. one particle state, to a superposition of ``plane waves'', i.e. superposition of single particle states, during the free-fall into the Schwarzschild BH, and vice versa by time symmetry. Furthermore, it should be noted that close to the horizon, according to eq~(\ref{bc1}), Eq.~(\ref{inverseTF}) is similar to Eq.~(19) in ~\cite{Solodukhin2004}
\begin{equation}
\Phi(u)= \int_{-\infty}^{+\infty}d\Omega\, \hat{\Phi}_0(\Omega)\, \Phi_{\Omega,0}(u)
\end{equation}
but, we can go one step further in this ``freely falling wave'' description. Indeed, in order to know the frequency-distribution of the locally accelerated ``wave'' $\Phi(u)$, we can Fourier transform it 
\begin{equation}\label{TF}
\hat{\Phi}_0(\Omega)=\int_{-\infty}^{+\infty}du'\, \Phi(u')\, e^{i\Omega u'}.
\end{equation}
We would like to note that what follows is not really ``rigorous'' in a quantum field theory sense, as it has been noticed in \cite{Alsing_Milonni2004}, but it permits us to have a physical intuition of what happens near the event horizon. According to eqs.~(\ref{horizon_state}) and (\ref{TF}) we can write explicitly
\begin{equation}\label{TFexplicit}
\hat{\Phi}_0(\Omega)=\int_{-\infty}^{+\infty}du'\, e^{i\Omega u'}\, \exp\left[-i\left(\frac{\omega}{\kappa}\right)e^{\kappa u'}\right].
\end{equation}
Thus, following \cite{Alsing_Milonni2004}, we change variables to $y=\exp(\kappa u')$ and write
\begin{equation}
\hat{\Phi}_0(\Omega)=\frac{1}{\kappa}\int_{0}^{+\infty}dy\, y^{i\Omega/\kappa-1}\, \exp\left[-i\left(\frac{\omega}{\kappa}\right)y\right].
\end{equation}
The integration reads
\begin{equation}
\hat{\Phi}_0(\Omega)=\frac{1}{\kappa}\Gamma\left(\frac{i\Omega}{\kappa}\right)\,\left(\frac{\omega}{\kappa}\right)^{i\Omega/\kappa}\exp\left(-\pi\Omega/2\kappa\right).
\end{equation}
The modulus squared of the amplitude of each ``plane waves'', i.e. $\left|\hat{\Phi}_0(\Omega)\right|^2$, is naturally interpreted as a probability (density) of measuring the associated frequency $\Omega$. Since we know from the properties of $\Gamma$ functions that 
\begin{equation}
\left|\Gamma\left(\frac{i\Omega}{\kappa}\right)\right|^2=\frac{\pi}{\left(\Omega/\kappa\right)\sinh\left(\pi\Omega/\kappa\right)}
\end{equation}
then the probability (density) of measuring the frequency $\Omega$ is given by a Bose-Einstein-like distribution
\begin{equation}
\left|\hat{\Phi}_0(\Omega)\right|^2=\left(\frac{2\pi}{\Omega\kappa}\right)\frac{1}{\exp\left(2\pi\Omega/\kappa\right)-1}.
\end{equation}
It is worth noting that this time dependent ``Doppler-gravitational'' shift effect can be adapted to fermions with the formal prescription $i\Omega/\kappa \rightarrow i\Omega/\kappa+1/2$. We refer the reader to \cite{Alsing_Milonni2004} for more details. Of course, the previous quantum point of view with superposition of particle states remains valid if one associates the frequency $\Omega$ with the energy of a quantum particle and interprets the probability of measuring $\Omega$ as the probability of finding particles of energy $\Omega$, as we will show it in section \ref{scattering_statistics}.

\section{The far horizon limit and the QNM complex frequencies}\label{HDQNM}

\subsection{The far horizon limit}
We start by using the transformation
\begin{equation}
\Phi_{\omega,\ell}(r)=\left(1-\frac{2M}{r}\right)^{1/2}\phi_{\omega,\ell}(r).
\end{equation}
Then, after a partial fraction decomposition, the Regge-Wheeler equation becomes
\begin{eqnarray}\label{precoulomb}
&&\frac{1}{4M^2}\frac{d^2\phi_{\omega,\ell}}{dr^2}+\left[\omega^2+\frac{J-3/4}{r^2}+\frac{J+\ell(\ell+1)-1/2}{2Mr}\right. \nonumber \\
&&\qquad \qquad \qquad \quad \left.+\frac{8M^2\omega^2-J-\ell(\ell+1)+1/2}{2M(r-2M)}+\frac{4M^2\omega^2+1/4}{(r-2M)^2}\right]\phi_{\omega,\ell}=0. \nonumber \\
&&
\end{eqnarray}
We substitute the variable $r$ by $x=(r-2M)/2M$ such that $x \in [0,+\infty[$ when $r \in [2M,+\infty[$. Thus, we consider large $x$ region, i.e. $r \gg 2M$, for which we can expand Eq.~(\ref{precoulomb}) in powers of $x^{-1}$
\begin{equation}\label{coulomb3}
\frac{d^2\phi_{\omega,\ell}}{dx^2}+\left[\omega^2+\frac{2\omega^2}{x}+\frac{\omega^2-\frac{\ell(\ell+1)}{4M^2}}{x^2}+\frac{1+\ell(\ell+1)-J}{4M^2x^3}+\ldots\right]\phi_{\omega,\ell}=0.
\end{equation}
It should be noted that in Eq.~(\ref{coulomb3}) the spin, $J=1-j^2$, only appears in higher orders of the expansion, starting from the third one. One way to solve such an equation is to consider the first two terms of the asymptotic expansion (\ref{coulomb3}) but the price to pay is the lost of information concerning the spin
\begin{equation}\label{coulomb}
\frac{d^2\phi_{\omega,\ell}}{dx^2}+\left[\omega^2+\frac{2\omega^2}{x}+\frac{\omega^2-\frac{\ell(\ell+1)}{4M^2}}{x^2}\right]\phi_{\omega,\ell}=0.
\end{equation}
Thus, the equation becomes analogous to a Coloumb differential equation which reads for an $s$-wave, i.e. $\ell=0$,
\begin{equation}\label{scoulomb}
\frac{d^2\phi_{\omega,0}}{dx^2}+\left[\omega^2+\frac{2\omega^2}{x}+\frac{\omega^2}{x^2}\right]\phi_{\omega,0}=0.
\end{equation}

\subsection{The $s$-wave scattering and the QNM complex frequencies}
In order to find the QNM complex frequencies, we will make some changes of variables. We introduce $b$ and $z$ such as
\begin{subequations}
\begin{eqnarray}
&&-4M^2\omega^2=b(b-1)\\ \label{bb}
&&-4iM\omega x=z. \label{zx}
\end{eqnarray}
\end{subequations}
Finally, we define a new ``$s$-wave function''
\begin{equation} 
\psi_{\omega}(z,b)=x^b \exp(2iM\omega x)\phi_{\omega,0}(z).
\end{equation}
Therefore, the equation (\ref{scoulomb}) reads
\begin{equation}\label{hypergeom}
z\frac{d^2 \psi_{\omega}}{dz^2}+(2b-z)\frac{d\psi_{\omega}}{dz}-(b-2iM\omega)\psi_{\omega}=0
\end{equation}
which is one of the well-known confluent hypergeometric equations \cite{AS65} (for the $\ell \neq 0$ case, we refer the reader to \cite{LiuMashhoon1996}). The solution of Eq.~(\ref{hypergeom}) is a combination of confluent hypergeometric functions of the first kind
\begin{eqnarray}
&&\psi_{\omega}(z,b)=\frac{\Gamma(1-2b)}{\Gamma(1-b-2iM\omega)}F_1\left(b-2iM\omega;2b;z\right)\nonumber \\
&& \qquad \qquad +\frac{\Gamma(2b-1)}{\Gamma(b-2iM\omega)}z^{1-2b}F_1\left(1-b-2iM\omega;2-2b;z\right).
\end{eqnarray}
Then the solution for the ``$s$-field'' $\phi_{\omega,0}$ reads
\begin{eqnarray}
&&\phi_{\omega,0}(x)=x^be^{2iM\omega x}\frac{\Gamma(1-2b)}{\Gamma(1-b-2iM\omega)}F_1\left(b-2iM\omega;2b;z(x)\right)\nonumber \\
&& \qquad \qquad +\left(-4iM\omega x\right)^{1-2b}x^be^{2iM\omega x}\frac{\Gamma(2b-1)}{\Gamma(b-2iM\omega)}z(x)^{1-2b}F_1\left(1-b-2iM\omega;2-2b;z(x)\right)\nonumber \\
&&
\end{eqnarray}
where $z(x)$ is defined by Eq.~(\ref{zx}).\\ 
For large frequencies, i.e. for $\omega>\omega_\text{min}$ where $\omega_\text{min}$ has been defined in section \ref{generalities}, we can use the following approximation
\begin{equation}
b_\pm \approx 1/2 \pm 2iM\omega \nonumber
\end{equation}
and, from the large $x$ limit, we use the asymptotic expansions of the previous confluent hypergeometric functions in terms of $\Gamma$ functions. Moreover, in order to obtain the amplitudes related to the ingoing wave, i.e. $A_{in}$, and to the outgoing wave, i.e. $A_{out}$, one has to note that $x$ and $r_\ast$ are related by the identity
\begin{equation}
e^{\pm2iM\omega x} x^{\pm2iM\omega}=e^{\pm i\omega r_\ast}. \nonumber
\end{equation}
Thus, up to numerical constants of order unity, the amplitude of the asymptotic behavior of the $s$-wave, for ``high frequencies'', reads
\begin{subequations}\label{AinAout}
\begin{eqnarray}
&&A_{in}(\omega)\approx (-4iM\omega)^{-b-2iM\omega}\frac{\Gamma(1-4iM\omega)\Gamma(4iM\omega)}{\sqrt{\pi}~\Gamma(1/2-4iM\omega)}\left(x^{-1+2b}b^{1-2b}+1\right)\\
&&A_{out}(\omega)\approx (4iM\omega)^{-b+2iM\omega}\frac{\Gamma(1-4iM\omega)\Gamma(4iM\omega)}{\pi}\left(x^{-1+2b}b^{1-2b}+1\right).
\end{eqnarray}
\end{subequations}
such as the ``$s$-field'' has an asymptotical behavior similar to Eq.~(\ref{bc2}). It should be noted that, in these expressions, after some simplifications, the value of $b_{+}=1/2+2iM\omega$ is the same for $A_{in}$ and for $A_{out}$. Then, as seen earlier, the QNM complex frequencies which are defined by $A_{in}=0$, i.e. by the poles of the function $\Gamma(1/2-4iM\omega)$, can be written as
\begin{equation}
\forall n \in \textbf{N},\qquad \frac{1}{2}-4iM\omega_n=-n 
\end{equation}
or, in other words
\begin{equation}
\forall n \in \textbf{N},\qquad 8\pi M\omega_n=2\pi i \left(n+\frac{1}{2}\right).
\end{equation}
Here, the highly damped QNM complex frequencies are obtained considering a ``far region'' limit, $r>2M$, and a ``high frequency'' regime, i.e. $\omega>\omega_\text{min}$. It should be noted that, in such calculation, $\omega_n$ has no real part, which is known to be non null and depending both on the spin $j$ of field and on the characteristics of considered BH. More particularly, one could think that not considering the third order term in the expansion in $x \gg 1$ in Eq.~(\ref{coulomb3}), is equivalent to consider the particular case of spin $j$ satifying
\begin{equation}
1-J=1-(1-j^2)=0.
\end{equation}
In this case, Eq.~(\ref{coulomb3}) would be equivalent to Eq.~(\ref{scoulomb}) if $j=0$. But we know that this ``$j=0$ result'' differs from the results discussed by Motl in \cite{Motl2002} obtained with the help of the powerful monodromy techniques. In other terms, the statement ``Eq.~(\ref{coulomb3}) would be equivalent to Eq.~(\ref{scoulomb}) if $j=0$'' seems to be wrong. Then, as we already noticed earlier, the exact expression of the ``spin $j$ dependent''-QNM frequency spectrum can't obviously be obtained by our second order approximation, i.e. Eq.~(\ref{scoulomb}) for which we only get the exact expression of the imaginary part.

\section{From scattering to statistics}\label{scattering_statistics}

\subsection{Quantum field theory: a very brief survey}

In quantum field theory \cite{ParkerToms2009}, in order to compute the spectrum of outgoing particles from a BH, one usually considers the coefficients of the Bogolubov transformation relating the well-known creation (resp. annihilation) operator $b_\Omega$ (resp. $b_\Omega^{\dagger}$) to $a_\omega$ and $a_\omega^{\dagger}$ such as
\begin{equation}
b_\Omega=\int d\omega \left(\alpha^{\ast}_{\Omega\omega}a_{\omega}-\beta^{\ast}_{\Omega\omega}a_{\omega}^{\dagger}\right)
\end{equation}
with the commutation relations
\begin{subequations}\label{commutation_relations}
\begin{eqnarray}
&&\left[\eta_{\omega_1},\eta_{\omega_2}^{\dagger}\right]=\delta\left(\omega_1-\omega_2\right)\\
&&\left[\eta_{\omega_1},\eta_{\omega_2}\right]=0=\left[\eta_{\omega_1}^{\dagger},\eta_{\omega_2}^{\dagger}\right]
\end{eqnarray}
\end{subequations}
where $\eta=a$ (resp. $b$) and $(\omega_1,\omega_2)=(\omega,\omega')$ (resp. $(\Omega,\Omega')$). The annihilation operators $b_\Omega^\dagger$ and $a_\omega^\dagger$ define respectively the Boulware vacuum state (in the $(u,v)$ coordinate system) and the Kruskal vacuum state (in the $(U,V)$ coordinate system) by
\begin{subequations}\label{vacuum_states}
\begin{eqnarray}
&&b_\Omega^\dagger \left|0_B\right\rangle=0 \qquad \text{Boulware vacuum}\\
&&a_\omega^\dagger \left|0_K\right\rangle=0 \qquad \text{Kruskal vacuum}.
\end{eqnarray}
\end{subequations}
It should be noted that using eq~(\ref{commutation_relations}), the normalization condition for the Bogolubov coefficients reads
\begin{equation}
\int d\omega \left(\alpha_{\Omega\omega}\alpha^{\ast}_{\Omega'\omega}-\beta_{\Omega\omega}\beta^{\ast}_{\Omega'\omega}\right)=\delta\left(\Omega-\Omega'\right).
\end{equation}
Then, from the standard mode expansions for the considered field operator both in the coordinates $(u,v)$ and $(U,V)$, and after some calculations, it follows that the coefficients of the Bogolubov transformation obey the well-known relation
\begin{equation}\label{exp_alphabeta}
\left|\alpha_{\Omega\omega}\right|^2=\exp\left(8\pi M\Omega\right)\left|\beta_{\Omega\omega}\right|^2.
\end{equation}
Thus, one can easily deduce the expectation value of the ``$b$-particle'' number operator, i.e. $N_\Omega=b_\Omega^{\dagger},b_\Omega$ in the Kruskal vacuum state and, more generally, the physics of the Hawking effect.

\subsection{The Hawking radiation: a scattering effect}
Let us focus, now, on the scattering/reflection coefficient of an ingoing $s$-waves, i.e. $S_0(\omega)=-R_0(\omega)$. We recall the main result (\ref{AinAout}) obtained in section \ref{HDQNM}, i.e.
\begin{subequations}
\begin{eqnarray}
&&A_{in}(\omega)\approx (-4iM\omega)^{-b-2iM\omega}\frac{\Gamma(1-4iM\omega)\Gamma(4iM\omega)}{\sqrt{\pi}~\Gamma(1/2-4iM\omega)}\left(x^{-1+2b}b^{1-2b}+1\right)\nonumber \\
&&A_{out}(\omega)\approx (4iM\omega)^{-b+2iM\omega}\frac{\Gamma(1-4iM\omega)\Gamma(4iM\omega)}{\pi}\left(x^{-1+2b}b^{1-2b}+1\right).\nonumber
\end{eqnarray}
\end{subequations}
Using Stirling's approximation, i.e. considering ``large frequencies'', i.e. $\omega>\omega_\text{min}$, one has
\begin{equation}
\Gamma(1/2-4iM\omega) \approx e^{4iM\omega}(4iM\omega)^{-4iM\omega}e^{-4\pi M\omega}
\end{equation}
then, we can easily deduce that the ``$\ell=0$'' reflection coefficient can be written as
\begin{equation}\label{R0}
R_0(\omega)=\frac{A_{out}(\omega)}{A_{in}(\omega)}\approx e^{4iM\omega}\times e^{-4\pi M\omega}.
\end{equation}
From Eq.~(\ref{R0}), the reflection probability off the ``$\ell=0$'' Regge-Wheeler potential barrier (cf. Fig.~\ref{0RWPot}) reads
\begin{equation}\label{Boltzmann_factor}
\left|R_0(\omega)\right|^2 = e^{-8\pi M\omega}.
\end{equation}
The equation (\ref{Boltzmann_factor}) could be interpreted as a Boltzmann factor characterized by a temperature $T=1/8\pi M$. With the help of Eq.~(\ref{exp_alphabeta}), we can now stress the link between the quantum field theory approach and the scattering approach, i.e.
\begin{equation}
\left|\alpha_{\Omega\omega}\right|^2=\left|R_0(\Omega)\right|^{-2}\, \left|\beta_{\Omega\omega}\right|^2.
\end{equation}

\subsection{Physics and thermodynamical aspects of the Schwarzschild BH}

In the following, we will associate every ``local plane wave'' to a ``single particle state''. According to the above results, the infalling motion of a quantum particle is analogous to a time dependent boost along the radial direction and consequently the proper distance of the infalling quantum particle exponentially decreases with time. Moreover, if the infalling particles lose enough energy, they can be considered, for an external observer, as ``trapped'' between the event horizon $r_h$ and the location of $V_{0,max}$, which then will play the role of a ``thermal atmosphere'' of the BH. Finally, as a consequence of the time dependent boost, this thermal atmosphere will become thinner as the particles eternally fall towards the horizon. We claim that this analysis could allow to clarify the link between conformal field approaches and BH scattering and could be at the origin of BH thermal effects. Indeed, an $s$-wave, with energy at least equals to $\omega_\text{min}$, can escape the BH without tunneling. Reciprocally, such $s$-wave will be able to penetrate the barrier from the outside and fall to the horizon (cf. Fig.~\ref{0RWPot}). It is worth noting that, in this case, the amplitude of the reflected wave off the ``$\ell=0$'' Regge-Wheeler potential barrier is exponentially small, but not null. Moreover, the mean energy of massless particles in thermal equilibrium at temperature $T=1/8\pi M$ is roughly of order of $T$ which is bigger than $\omega_\text{min}$. Therefore, some of the $s$-waves, i.e. ``$s$-states'' particles, will easily escape to infinity. Unless the BH is kept in equilibrium by incoming radiation, it will lose energy to its surroundings. In other words, the BH evaporates. This is one possible explanation of the Hawking radiation in terms of $s$-waves scattering, without tunneling. It should be noted that it is not the case for fields (or particles) with $\ell \gg 1$ because the Regge-Wheeler potential barrier becomes large enough to reduce significantly this process. Moreover, in this case, the location of $V_{0,\text{max}}$ moves away from the horizon to be closer to the location of the photon sphere which will play a central role in the analysis of weakly damped QNM \cite{YDAFBR2010}. 
\newline
Furthermore, up to a normalizing contant, the expression (\ref{Boltzmann_factor}) tells us that the probability of finding one particle reflected by the ``$\ell=0$'' Regge-Wheeler potential barrier is the same as the probability of finding one particle of energy $\omega$ in a system, i.e. the thermal atmosphere, in thermodynamical equilibrium at temperature $T=1/8\pi M$. Thus, the probability $P_R(N_k)$ to find $N_k$ particles of energy $\omega_k$ in the thermal atmosphere of the BH, in thermodynamical equilibrium at the temperaure $T=1/8\pi M$ is
\begin{equation}
P_R(N_k)=\frac{1}{Z}\left(\left|R_0(\omega_k)\right|^2\right)^{N_k}=\frac{1}{Z}e^{-8\pi M\omega_k N_k}
\end{equation}
where the normalizing constant $Z$ is the partition function such as the probabilities sum up to one 
\begin{equation}
Z=\frac{1}{\sum_{N_k} P_R(N_k)}.
\end{equation}
The mean number of reflected particles depends of course on the nature of the considered particles and is usually given by
\begin{equation}
\left<N^{(F/B)}(\omega_k)\right>=\sum_{N_k=0}^{1\, \text{or} \,\infty}N_k\, P_R(N_k)
\end{equation}
where $F/B$ stands for ``Fermion/Bosons''. More explicitly
\begin{subequations}
\begin{eqnarray}
&&\left<N^{(F)}(\omega_k)\right>=\sum_{N_k=0}^{1}N_k\, P_R(N_k)=\frac{1}{e^{8\pi M\omega_k}+1} \quad \text{for\, fermions}\\
&&\left<N^{(B)}(\omega_k)\right>=\sum_{N_k=0}^{\infty}N_k\, P_R(N_k)=\frac{1}{e^{8\pi M\omega_k}-1} \quad \text{for\, bosons}.
\end{eqnarray}
\end{subequations}
From the Fourier analysis of section \ref{NHL}, one then has
\begin{equation}
\left<N^{(F/B)}(\Omega)\right> \propto \left|\hat{\Phi}_0^{(F/B)}(\Omega)\right|^2.
\end{equation}
Therefore, the mean number of particles of energy $\Omega$ is proportional to the probability of measuring a frequency $\Omega$ in the spectrum of the field $\Phi(u)$.

\section{Statistical ``heuristics''}\label{statistics}
As we have just seen, in thermodynamical terms, the system in thermal equilibrium at temperature $T=1/8\pi M$ is the thin thermal atmosphere located close to the event horizon. In this sense, it is tempting to think that the Physics which has to be considered to describe the spectrum of the Hawking radiation would be a conformal field theory, in the near vicinity of the horizon, associated to the field in interaction with the considered BH. Therefore, due to the accumulation of all the Fourier components (quantum particles states) of the incoming wave spectrum on this thermal layer, it would not be a surprise to describe the near horizon physics in terms of some ``other/new'' degrees of freedom which would then be expected to have rather ``exotic'' statistics. In the following, we will give a heuristic point of view concerning these new statistics.

\subsection{A naive approach}
Let us introduce the usual variable $\beta=T^{-1}$ to simplify the notations and let us consider ``$\ell=0$'' fermions reflected off the Regge-Wheeler potential barrier. The mean number of fermions with energy $\omega_n$ is given by the well-known Fermi-Dirac statistics
\begin{equation}
\left<N_i(\omega_n)\right>=\frac{1}{e^{\beta \omega_n}+1}=\frac{\left<N_i\right>}{g_i}
\end{equation}
where $g_i$ is the degeneracy of the $i^{th}$ state of energy $\omega_n$.
The QNM complex frequencies can be seen as poles in the complex $\omega$-plane of the Fermi-Dirac distribution, for which one can write
\begin{equation}
e^{\beta \omega_n/T}+1 \approx 0 \Rightarrow \beta \omega_n \approx 2\pi i\left(n+\frac{1}{2}\right)
\end{equation}
which is the result obtain in the previous section.\\
Even if the behavior is correct for large $n$, this approach does not give, obviously, the correct answer with the well-known constant real part, i.e. $\ln(3)/8\pi M$. How can one have access to the latter? One way to answer this question is to assume that near the horizon, there may be another, ``exotic'' statistics satisfied by the considered quanta which would not behave neither like fermions or bosons.\\
A first very naive approach is to consider a non-null chemical potential $\mu$, or a corresponding fugacity $z=e^{\mu/T}$, for a thermodynamical system with a conserved number of fermions. Then the Fermi-Dirac distribution reads
\begin{equation}
\left<N_i(\omega_n-\mu)\right>=\frac{1}{z e^{\beta \omega_n}+1}=\frac{z^{-1}}{e^{\beta \omega_n}+z^{-1}}.
\end{equation}
One could consider the fugacity $z$ as an universal degeneracy $g=z^{-1}$ for each state of an equivalent thermodynamical problem where the number of particles is now not conserved and, by definition, which statistics would be given by
\begin{equation}
\left<\tilde{N}_i(\omega_n)\right> = \frac{1}{g} \left<N_i(\omega_n-\mu)\right> = \frac{1}{e^{\beta \omega_n}+g}.
\end{equation}
Then, the QNM complex frequencies, seen as the poles of the statistical distribution are
\begin{equation}
e^{\beta \omega_n}+g \approx 0 \Rightarrow \beta \omega_n \approx \ln\left|g\right|+ 2\pi i\left(n+\frac{1}{2}\right).
\end{equation}
The exact analytical result suggest that for a Schwwarzschild BH, $g=3$. In other words, each energy state could have 3 possible sub-levels. We are conscious that such explanation is just a mathematical trick and that the underlying physics is still expected.

\subsection{More on statistics: the fractional statistics}

However, one could assume that near the event horizon of the Schwarzschild BH, particles behave neither like bosons or fermions but they satisfy a more general statistics which could be the Polychronakos' fractional statistics \cite{Polychronakos1996}, the Haldane's fractional statistics \cite{Haldane1991} or even the infinite statistics \cite{AltherrGrandou1993}, introduced some years ago, which all start by a generalization of the Pauli exclusion principle (highly suggested by the Motl's ``tripled Pauli statistics'' \cite{Motl2002}). In the former case, one could use the simple Eq.~(22) of \cite{Polychronakos1996} which reads
\begin{equation}
\left<N_i(\omega_n,\alpha)\right>=\frac{1}{e^{\beta \omega_k}+\alpha}.
\end{equation}
Fermions and bosons corresponds respectively to $\alpha=1$ and $\alpha=-1$. If we introduce the variable $g$ such as $\alpha=2g+1$, then
\begin{equation}
\left<N_i(\omega_n,g)\right>=\frac{1}{e^{\beta \omega_n}+1+2g}
\end{equation}
where fermions and bosons corresponds now respectively to $g=0$ and $g=-1$. We deduce naturally
\begin{equation}
\beta \omega_n=\ln\left|1+2g\right|+2\pi i \left(n+\frac{1}{2}\right).
\end{equation}
One can noticed that the choice $g=1$ ($\alpha=3$) allows us to recover the intriguing ``tripled Pauli statistics''. Let us note that the different expressions of highly damped QNM complex frequencies can be seen as statistics obtained for different kind of BH which could be embedded in the ``richness'' of these statistics \cite{Polychronakos1996,Haldane1991,AltherrGrandou1993}. In particular, we think that the fractional statistics could be compatible with the ``general'' expression defining the quasinormal frequencies given in \cite{Skakala2012}.

\section{Conclusion}
In our scattering picture, the Hawking radiation and thermal effects for a Schwarzschild BH are associated with the scattering of the $s$-waves by the ``$\ell=0$'' Regge-Wheeler potential barrier. A very interesting result is that the scattering/reflection amplitude is exponentially small but not null and is exactly the Boltzmann factor related to Hawking effect. We have seen, from an external observer perspective, that incoming waves (quantum particles) can be ``trapped'' between the location of $V_{0,\text{max}}$ and the horizon. Moreover, this ``thermal atmosphere'' becomes thinner as a consequence of the time dependent boost, i.e. the free-fall of the waves (quantum particles) into the BH. Then one could ask: is there a thickness limit for the ``thermal atmosphere''? According to the well-known thermodynamical results, the answer would probably be linked to the existence of a cutoff, maybe the Planck length, but this is way too far from the scope of our analysis. Furthermore, we claim that this thin thermal atmosphere could be at the origin of new exotic statistics, due to the change of degrees of freedom during the transition between the incoming field in spacetime and the accumulation of all its Fourier components (quantum particles states) on the ``thermal layer''. It is worth noting that this thermodynamics, and consequently all the associated quantities, is usually associated to the BH, seen as one thermodynamical system. We have thus two different origins for the highly damped QNM complex frequencies. The first is the damping of the QNM due to the scattering/reflection off the ``$\ell=0$'' Regge-Wheeler potential barrier in the ``far region limit'', directly described by the imaginary part. The second would be the thermodynamics of the ``thermal layer'' associated with the real part. Our scattering analysis seems to be in agreement with the Maggiore's interpretation \cite{Maggiore2008}. It should be noted that the Hawking effect and the highly damped QNM are not linked to a very low frequency limit \cite{Matzner1968} but simply to the interaction of $s$-fields with the considered BH, whatever their energy is. Moreover, in addition to the existence of a thermal atmosphere, we have also recovered some results obtained in \cite{Solodukhin2004} but through conformal field frameworks, suggesting naturally the use of such approaches to describe the physics near the event horizon. Furthermore, one has to bear in mind that the statistics of particles should be thought as a property of spacetime and particularly near the horizon of a Schwarzschild BH. Therefore, if spacetime is quantum by nature, the commutation or anticommutation relations for creation and annihilation operators, for the considered fields, may change \cite{Swain2009}. Thus, exotic statistics and thereby a modification of the spin statistics theorem could be deeply related to the very quantum nature of spacetime and could emerge from an underlying theory of quantum gravity in the near vicinity of a BH horizon. Finally, it is worth noting that a generalization to $d$-dimensional Schwarzschild-Tangherlini BH in interaction with a scalar ($j=0$) field is trivial, at least for the imaginary part of the QNM complex frequencies. Indeed, in this case, the ``dimension parameter'' in the Regge-Wheeler potential is multiplied by the angular momentum $\ell$ (for more details, see \cite{YDAFBR2010}). Thus, for the scattering of an $s$-wave, the results will be identical. Concerning the real part, there is no ``simple'' argument except that the ``$\ell=0$'' Regge-Wheeler potential is not modified and thus one could perform the same calculations and recover the same scattering/reflection amplitude as in section \ref{HDQNM}. Actually, the real part seems to not be affected either by the dimension of spacetime as it has been noticed by Motl in \cite{Motl2002} (and references therein).

\acknowledgments
I would like to thank Professors Antoine Folacci, Yves Decanini, Jean-Pierre Provost, Christian Bracco, Thierry Grandou and Sergey Solodukhin for very stimulating discussions and support.

\bibliography{Biblio}

\end{document}